\documentclass[pra,twocolumn]{revtex4}

\usepackage{graphicx}
\usepackage{amssymb}
\usepackage{amsmath}
\usepackage{amsbsy}
\usepackage{xcolor}
\usepackage{upgreek}
\usepackage{siunitx}
\usepackage{tikz}
\usepackage{soul}

\usepackage{lipsum}

\usepackage{multirow}

\usepackage{hyperref}

\hypersetup{
    colorlinks=true,
    linkcolor=blue,
    filecolor=magenta,      
    urlcolor=blue,
    citecolor=red
}
 
\begin{document}

\author{\bf Eric Giglio}

\title{What is wrong with the image charge force of keV ions in insulating nano-capillaries ?}
\date{\today}
\affiliation{Centre de Recherche sur les Ions, les Mat\'eriaux et la Photonique (CIMAP), Normandie Univ, ENSICAEN, UNICAEN, CEA, CNRS, F-14000 Caen, France}

\pacs{34.80.Dp, 34.80.Pa}

\begin{abstract}

%When an ion approaches a dielectric interface separating two different dielectric media, the induced surface polarization charges attract the ion toward the interface. 
In nano-capillaries of large aspect ratio, 
the attractive image charge force is strong enough to affect the trajectory of  ions passing through capillaries and consequently to diminish the fraction of transmitted beam ions. 
We calculated the theoretically transmitted  fraction, 
using an approached but CPU-friendly expression of the image charge force valid in the case of a static ion and an infinite cylindrical dielectric interface. When comparing the theoretically transmitted fraction to  available experimental data for nano-capillaries with an inner diameter of less than 200 nm, we found a surprisingly large disagreement, i.e.,   the theoretically transmitted fractions   were easily an order of magnitude lower than the experimental ones.
Noting that  the image charge force depends on the velocity of the ion via the frequency dependent relative permittivity of the insulator, we investigated whether the disagreement could be lifted using a velocity depend image charge force.
We give the exact expressions of the dynamical image charge force for a plane and cylindrical dielectric interface as a function of the ion velocity. We then re-evaluated the theoretically transmitted fractions in the case of  SiO$_2$ and PET dielectric interfaces. Our findings are discussed in the light of the available experimental data.
\end{abstract}

\maketitle

\section{Introduction}

Guiding of low-energy ions by insulating capillaries
was first reported by Stolterfoht {\it et al.} in 2002 for nano-capillaries \cite{Stolterfoht_2002} and later by Ikeda {\it et al.} \cite{Ikeda} for macroscopic glass capillaries. Their experimental results 
showed that even though the capillaries were tilted with respect to the beam axis such as no geometrical transmission was allowed, the beam was steered after  an initial charge-up phase  through the insulating capillary by self organized charge patches. As a result, a part of the beam could be transmitted, with the ions keeping their initial charge state, indicating that the ions never touched the inner wall. Those pioneering results triggered numerous experimental and theoretical studies to better understand the guiding process of keV ions through insulating capillaries \cite{Schiessl05,Sahana_PRA_2006}. For an overview, the reader may consider \cite{Lemell_2013} and \cite{Stolterfoht_2016}.  By combining available experimental data with theoretical predictions, insulating capillaries become a formidable tool to explore the  dynamics of excess charges in insulators in the presence or absence of charge injection. Indeed, the time-evolution of the rate, energy and angular profile of the transmitted ions give indirectly information about the charge injection process by ion-surface collisions  and about the relaxation process of excess charge in insulating capillaries. 

On the theoretical front, models for the charge dynamics in insulating capillaries were contentiously refined over time in order to improve the description of the charge injection and charge relaxation  in insulating capillaries \cite{Schiessl07,Stolterfoht13-1, Giglio_PRA_2018, GIGLIO_2020, Niko_atoms_2020, GIGLIO_PRA_2021}. In particular, the electric field inside the capillary was calculated precisely by considering appropriate boundary conditions at the inner and outer surface of the capillary \cite{GIGLIO_2020}. In recent models for glass macro-capillaries,  the charge injection at the inner surface term accounts even for secondary electrons emitted by ion-surface collisions and subsequently re-absorbed by the surface \cite{GIGLIO_PRA_2021}.  In nano-capillaries, the bulk and surface currents were even considered to depend  non-linearly on the electric field as the electric field in nano-capillaries exceeds easily the keV/mm \cite{Niko_atoms_2020}. Eventually, the image  charge force acting on the beam ions inside nano-capillaries  was added to the dynamics   to explain  the observed beam-shaping of  transmitted ions through  muscovite mica nano-capillaries with rhombic  and rectangular section \cite{Zhang_2012,Zhang_2017}.  Indeed, when an ion approaches a dielectric interface separating two different dielectric media, the induced surface polarization charges attract the ion toward the interface. 
In nano-capillaries, the attractive image charge force was shown to be  strong enough to  sensibly affect the trajectory of  beam ions  and consequently the fraction and angular distribution of transmitted beam ions.

%Eventually, as suggested by experimental results,  the image  charge force acting on the beam ions inside the capillary is taken into account \cite{Zhang_2012}. All this theoretical effort lead to simulations that not only produced results that compared well to experimental data but and may be used to give reliable numerical predictions that can be verified experimentally  \cite{Giglio}. 

In this work, the author proposes to focus in more detail on the modeling of  the image charge force acting on beam ions through cylindrical nano-capillaries. In particular, we want to investigate whether  experimentally observed  transmission rates through insulating straight nano-capillaries are well reproduced using the expression of the image force inside a cylindrical dielectric. Several experimental studies reported the transmission rates of keV ions through insulating nano-capillaries \cite{Sahana_PRA_2006,Niko_PRA_2010,Niko_EPJ_2021,Zhang_2017}. Here, we are mostly interested in those cases where a uniform keV ion beam was injected at zero tilt angle and for which the measured transmitted fraction, defined as the transmitted current divided by the injected current, was given. These conditions constitute an ideal setup for which the transmitted fraction can be calculated analytically as the ion  trajectories are dominated by the image charge force.
Indeed, as will be discussed further in the text, for a  uniform beam, aligned with the capillary axis, the charge deposited in the capillary can be assumed axisymmetric, yielding an inner electric field  that is  negligible compared to the image charge force. 
As a result, ion beams  through zero-tilted nano-capillaries  are a formidable tool  to investigate  the image  force of ions moving along a cylindrical dielectric interface.

The paper is organized as follow. In section \ref{sec_disagreement}, we present and discuss  a  CPU-friendly expression of the image force acting on a  charge inside a cylindrical dielectric interface. We then calculate   the theoretical transmitted fraction in the case where the injected beam is aligned with the capillary axis. Next, we highlight the discrepancy between the theoretical and the experimental transmitted fractions found in the literature. That finding made us question the validity of the expression of the image charge force used here. In section \ref{sect_model_imforce} we propose to improve the expression of the image charge force by explicitly  taking into account the velocity of the ion and the frequency dependence of the relative permittivity of the capillary. We then calculate the image charge force as a function of the ion velocity and discussed its influence in the case of an arbitrary dielectric response function. 
In section \ref{sec_numerics} we re-evaluated the theoretical transmitted fractions using this time the velocity depend image charge force, mainly for those cases for which experimental data is available in the literature. We discuss then in how far the velocity dependent image charge force lifts the initially mentioned  discrepancy.

\section{Theoretical and experimental transmitted fractions}

\label{sec_disagreement}

In the following, we  highlight  the apparent disagreement between the theoretical and experimentally observed  transmitted fraction and of ions trough nano-capillaries. In a first step we will define the theoretically transmitted fraction $f_\text{th}$ of an injected ion beam at zero tilt angle, in the ideal case where the capillary is a straight tube and the uniform, mono-kinetic ion beam has zero divergence.  Then, we compare the theoretical transmitted fraction to the one found in former published works and discuss the observed discrepancies. 

\begin{figure}[h!]
	\includegraphics[scale=0.39]{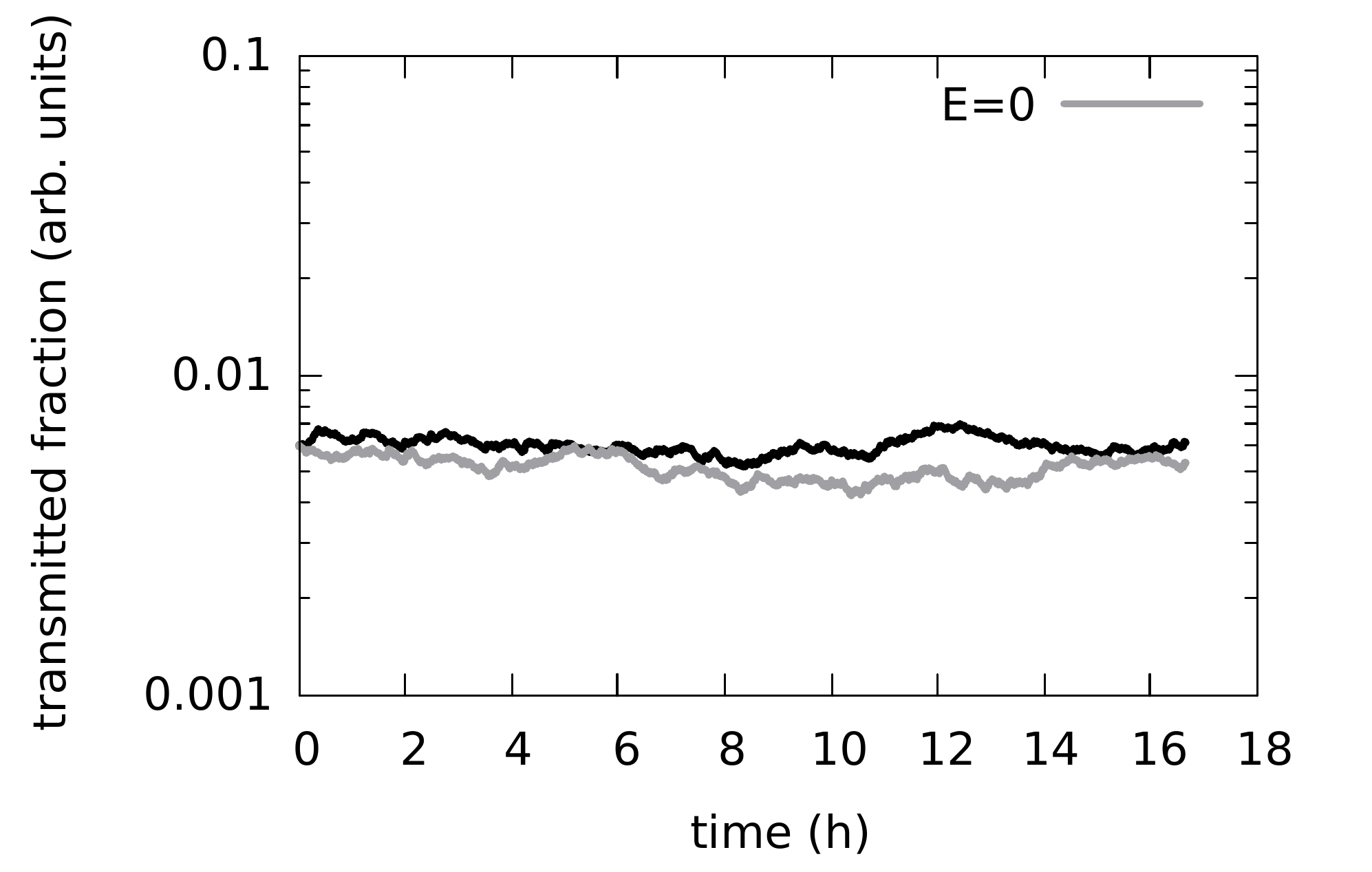}
	\caption{Simulated transmitted beam fractions (black curve) through an insulating nano-capillary. The gray curve,  for which the electric filed due to deposited charges was removed, has been added for comparison.   
\label{fig_trans_sio2}}
\end{figure}

The ion trajectories through charged  capillaries is mainly affected by two fields, the electric field generated by the accumulated charges in the capillary walls and  by the image force. For a beam aligned with the capillary axis the deposited charge distribution has axial symmetry,  generating an axisymmetric  electric potential $V(\rho,z)$. Such a potential yields an inner electric field that competes with the image force only if the accumulated charge density in the bulk exceeds a certain amount. But because the insulator  cannot withstand an electric field larger than the dielectric strength field, typically $50$ kV/mm \cite{Dash_2018},  without undergoing electrical breakdown and becoming electrically conductive, the accumulated charge  in the capillary wall is limited. We performed simulations with our numerical code InCa4D \cite{Giglio_PRA_2018} where we followed the trajectories of keV ions through  a SiO$_2$ nano-capillary.  We monitored  the ion transmission for several hours ( see black curve in figure \ref{fig_trans_sio2}) and found that the transmission is stable in time and neither increased or reduced by the deposited self-organized charge. Details of the calculations are presented in Appendix \ref{appendix_radial_electric}. Simulations confirm that the effect of the deposited charge on the transmission is negligible and that it is not possible to charge an insulating nano-capillary to the necessary amount to generate an inner electric field that competes with the image force. 
%
%   Using $\nabla^2 V=0  $, the radial component of the inner electric  field in a large aspect capillary is well approximated by 
%\begin{equation}
%E_\rho(\rho) \simeq \frac{1}{2}\rho \left.\frac{\partial^2 V}{\partial z^2}\right|_{\rho=0}
%\end{equation}
%with 
%\begin{equation}
%V(0,z)=\frac{1}{2 \varepsilon_0}\int \frac{\sigma(z') R dz'}{\sqrt{R^2+(z-z')^2}}
%\end{equation}
%For a charge patch of length $h$ and surface density $\sigma_0$, $V''$ scales as $\sigma_0 R h^{-2}/\varepsilon_0 $. Now, in dielectrics the electric strength defines the maximal electric field an insulator can withhold before breaking up. 
%
% becasue the deposiyed charges  we can 
%
%Because the injected ion beam can deposit charges into the capillary walls, we must check whether the electric field generated by the deposited charges can be considered negligible compared to the image force.  an ion beam is injected into the capillaries, Note that the radial electric field to  The deposited  charge, controlled by the radial image force  
%
%has thus axial symmetry and for nano-capillaries of large aspect ratio, the electric field due to the deposited charges is typically negligible compared to the image force. Indeed the electric field generated by the deposited charges inside nano-capillaries becomes non-negligible only for an amount of  deposited charges that would generate an electric field in the capillary walls that exceeds by 3 orders of magnitude the typical 
%electric strength of the dielectric, see appendix \ref{appendix_radial_electric}.  
We assume thus in the following that the theoretically transmitted fraction depends  only on the image force that acts on the ions passing through the capillaries.
\subsection{Theoretically transmitted fraction}
\label{sec_theo_trans_frac}

\subsubsection{Image charge force at a cylindrical dielectric interface}

Let us assume a translationally invariant cylindrical dielectric interface of radius $R$, separating the inner vacuum from the outer insulating medium of dielectric relative permittivity $\varepsilon_r>1$. A charge $q$ is located inside the cylinder at position $\rho< R$, where $\rho$ is the distance from the symmetry axis Oz. The modeling of  the image  force acting on the charge $q$ is described in details in \cite{Karoly_NIMB_1999,Karoly_PRA_2001} and  \cite{CUI_2006}.   The resulting force, expressed in cylindrical coordinates ($\rho,\theta,z$) is purely radial and given as a sum over angular moments $m$ and wave-numbers $k$, 
\begin{equation}
\vec{F}_\text{im}(\rho) = \frac{q^2}{16 \pi \epsilon_0 } \sum_{m=0}^\infty \int_0^\infty \!  \!  \!  \!dk \, A_m(k R,\varepsilon_r)  \frac{\partial I^2_m(k\rho)}{\partial \rho}  \vec{u}_\rho \;,
\label{eq_im_force_exact}
\end{equation}
where  $\vec{u}_\rho$ is the radial unit vector, $I_m()$ the modified Bessel function of order $m$ and  $A_m()$ a dimensionless function, accounting for the dielectric permittivity $\epsilon_r$ of the capillary.  The expression of $A_m()$ is given by Eq. \ref{eq_chi_1}.
Because of the sum over $m$ and $k$, computing the image charge force $F_\text{im}$ may be too time consuming in simulations and a simpler, CPU-friendly expression approaching $F_\text{im}$ would be welcomed.  
On the basis of the image charge force of a  perfectly conducting cylindrical interface (Eq. \ref{eq_imforce_conduct}) given by K\"ok\'esi \textit{et al.} \cite{Karoly_NIMB_1999,Karoly_PRA_2001},
we proposed to approach  $ \vec{F}_\text{im}$  by the following CPU-friendly expression,
\begin{equation}
\vec{F}^{0}_\text{im}(\rho) = \frac{q^2}{4 \pi \epsilon_0 } \frac{\epsilon_r-1}{\epsilon_r +1} \frac{\rho R}{(R^2-\rho^2)^2} \vec{u}_\rho \quad.
\label{eq_im_force}
\end{equation}

\begin{figure}[h!]
	\includegraphics[scale=0.28]{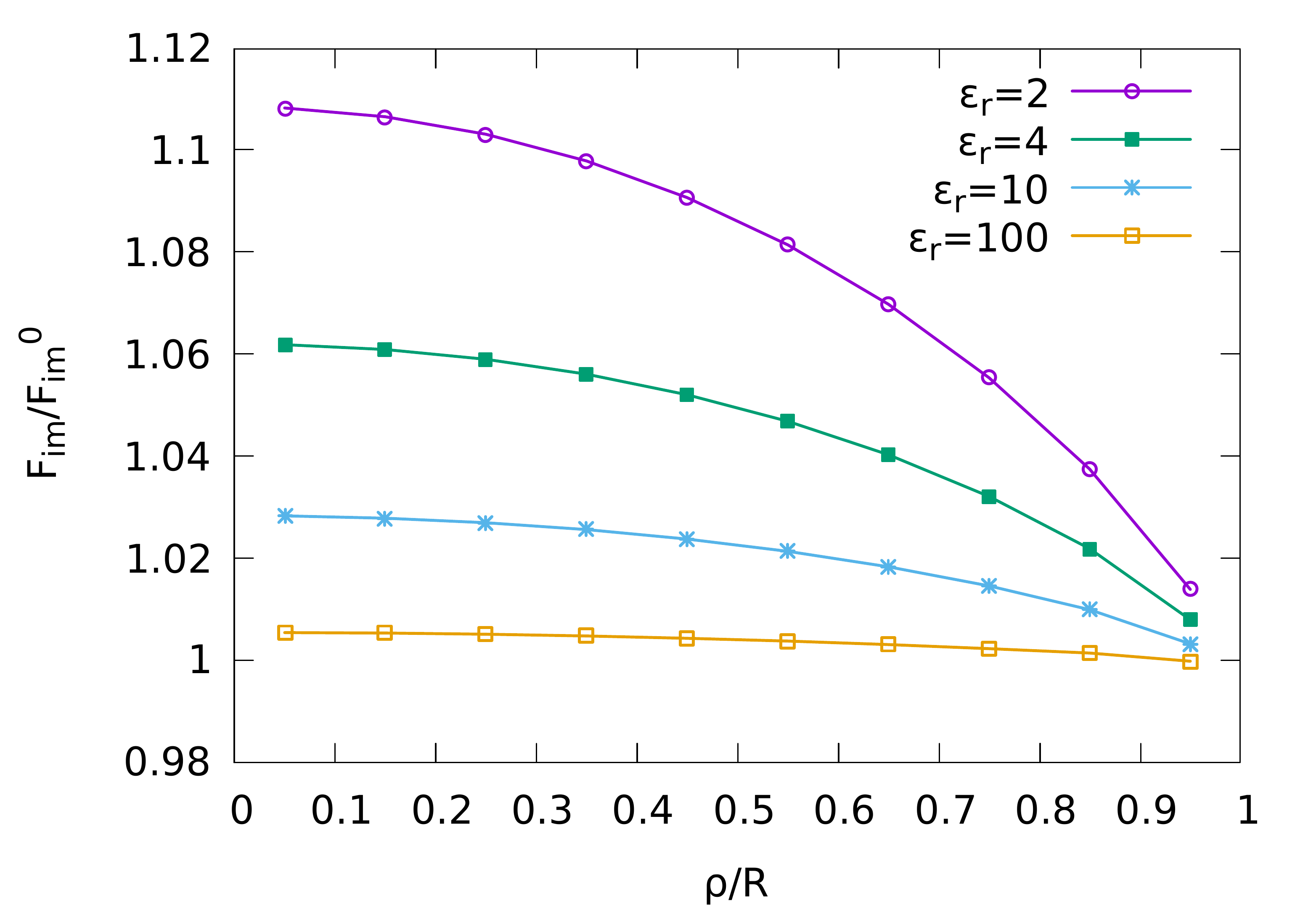}
	\caption{Image charge force  $F_\text{im}$ (Eq. \ref{eq_im_force_exact}) divided by the force $F_\text{im}^{0}$  (Eq. \ref{eq_im_force}) as a function of the ratio $r=\rho/R$ for various dielectric constants $\varepsilon_r$
\label{fig_imforce}}
\end{figure}

In order to check how well $ F^{0}_\text{im}$ approaches $F_\text{im}$, we compute the ratio  $F_\text{im} / F_\text{im}^{0}$ as a function of $\rho/R$ and for various values of $\varepsilon_r$. Results are shown in Fig. \ref{fig_imforce}.   $F_\text{im}(\rho)$ was evaluated using $R=100$ nm and the sum over $m$  was limited to the first 100 angular moments, which was found to be sufficient for an accurate evaluation of (\ref{eq_im_force_exact}) up to a radial position $\rho/R \le 0.98$. The relative error decreases with increasing $\varepsilon_r$, but even for $\varepsilon_r=2$,  the ratio is close to unit, with a relative error of about than 10\%, meaning that the exact expression  $F_\text{im}$ is well approached by $ F_\text{im}^{0}$. In particular, close to the interface, $\rho/R \rightarrow 1 $, $ F_\text{im}^{0}$ tends asymptotically to $ F_\text{im}$. In our simulations and for the evaluation of the theoretical transmitted fraction, we will thus use $ F_\text{im}^{0}(\rho)$ instead of the exact but CPU consuming expression  (\ref{eq_im_force_exact}) to model the image charge force that acts on the ion inside the cylinder.  

\subsubsection{Ion trajectories and the theoretical transmitted  fraction} 
\label{sec_trajectories}
We assume that a charge $q$ of mass $m$ and energy $E_k$ enters the cylindrical capillary with an initial radial distance $\rho_0<R$  and  with an initial velocity  $\vec{v}_0=\sqrt{2 E_k/m} _,\vec{u}_z$, so that at $z=0$ one has $d\rho/dz=0$. The trajectory $\rho(z)$ of the ion can then be readily computed by solving the differential equation
\begin{equation}
\frac{d^2 \rho}{d z^2} = \frac{1}{2 E_k} F^{0}_\text{im}(\rho(z))
\label{eq_traj}
\end{equation}
with the initial values $\rho(0)=\rho_0$ and $\left.\frac{d \rho}{d z}\right|_{z=0}=0$.  
Introducing the dimensionless  quantities $\varrho=\rho/R$ and 
$\zeta=z/H$, 
equation (\ref{eq_traj}) can be re-written in dimensionless form
\begin{eqnarray}
\frac{d^2 \varrho}{d \zeta^2} &=& \Gamma \frac{\varrho(\zeta)}{(1-\varrho(\zeta)^2)^2}  
\label{eq_traj_dimless}
\end{eqnarray}
with $ \varrho_0=\rho_0/R$ and $\left. \frac{d\varrho}{d\zeta}\right|_{\zeta=0} = 0$ and where the dimensionless factor
\begin{equation}
\Gamma =\frac{q^2}{4 \pi \varepsilon_0} \frac{\varepsilon_r-1}{\varepsilon_r +1} \frac{1}{2 E_k} \frac{H^2}{ R^3}
\label{eq_k}
\end{equation}
accounts for the length $H$, inner radius $ R$ and dielectric constant $\varepsilon_r$ of the capillary as well as the kinetic energy $E_k$ of the charge. Remarkably, the factor $\Gamma$ tells us that the dimensionless image charge force scales with the square of the capillary length $H$ and but with  the inverse cube of the capillary radius $R$. This explains why, for nano- and macro-capillaries of same aspect ratio $H/R$, the force can be neglected in macro-capillaries ($R\sim 1$ mm)  but not in nano-capilleries ($R\sim 100$ nm). Also note that the force is inversely proportional to the kinetic energy of the ions. 

\begin{figure}[t!]
	\includegraphics[scale=0.36]{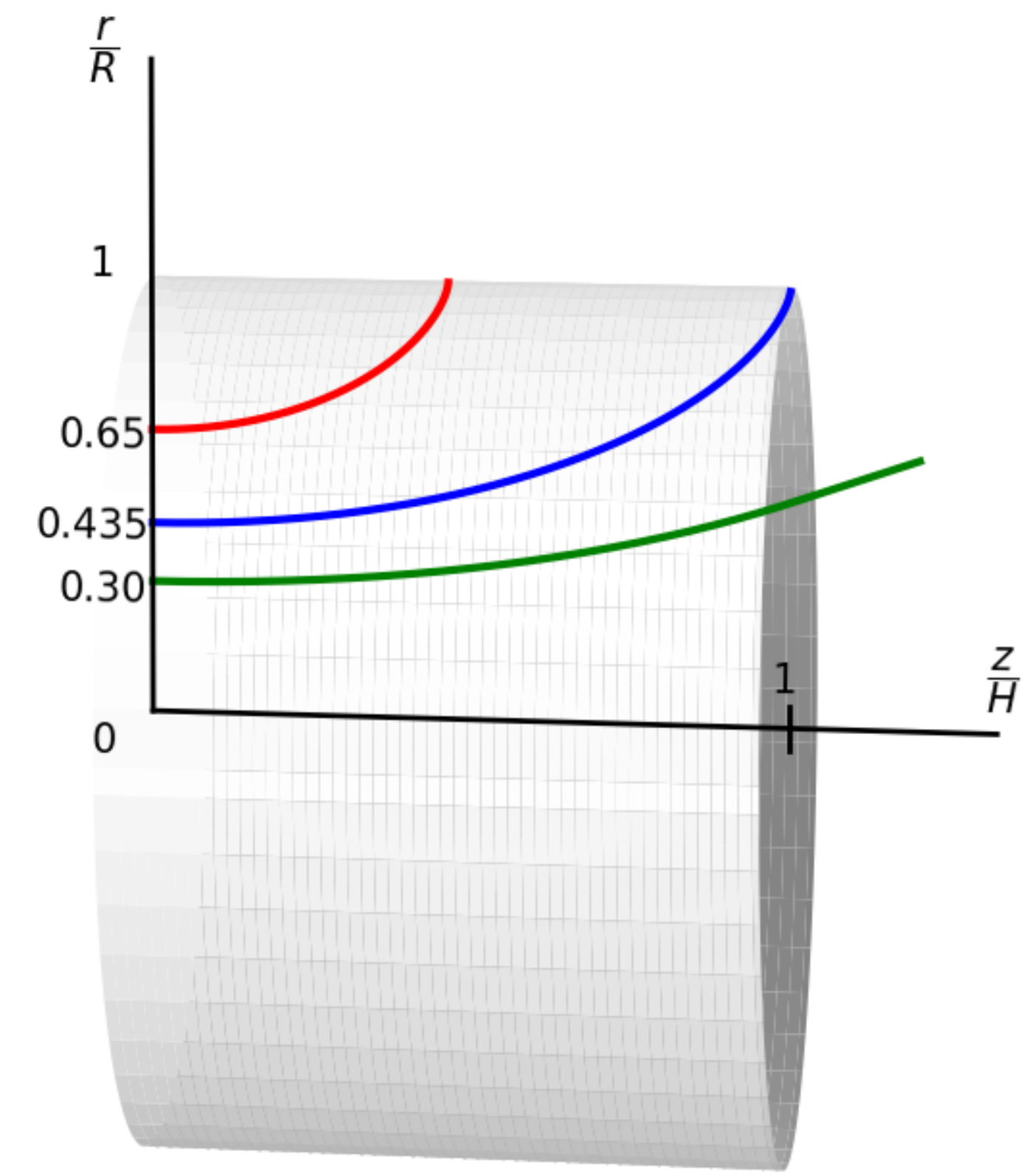}
	\caption{Trajectories for 3 different initial values of $\varrho_0$ in the case of $\Gamma=1$. Only for $\varrho_0^{\text{max}}<0.435$ are the ions transmitted, so that for $\Gamma=1$ the transmitted fraction $f_\text{th}$ equals $\varrho^2=18.6$\% 
\label{fig_traj_rho}}
\end{figure}

\begin{figure}[h!]
	\includegraphics[scale=0.4]{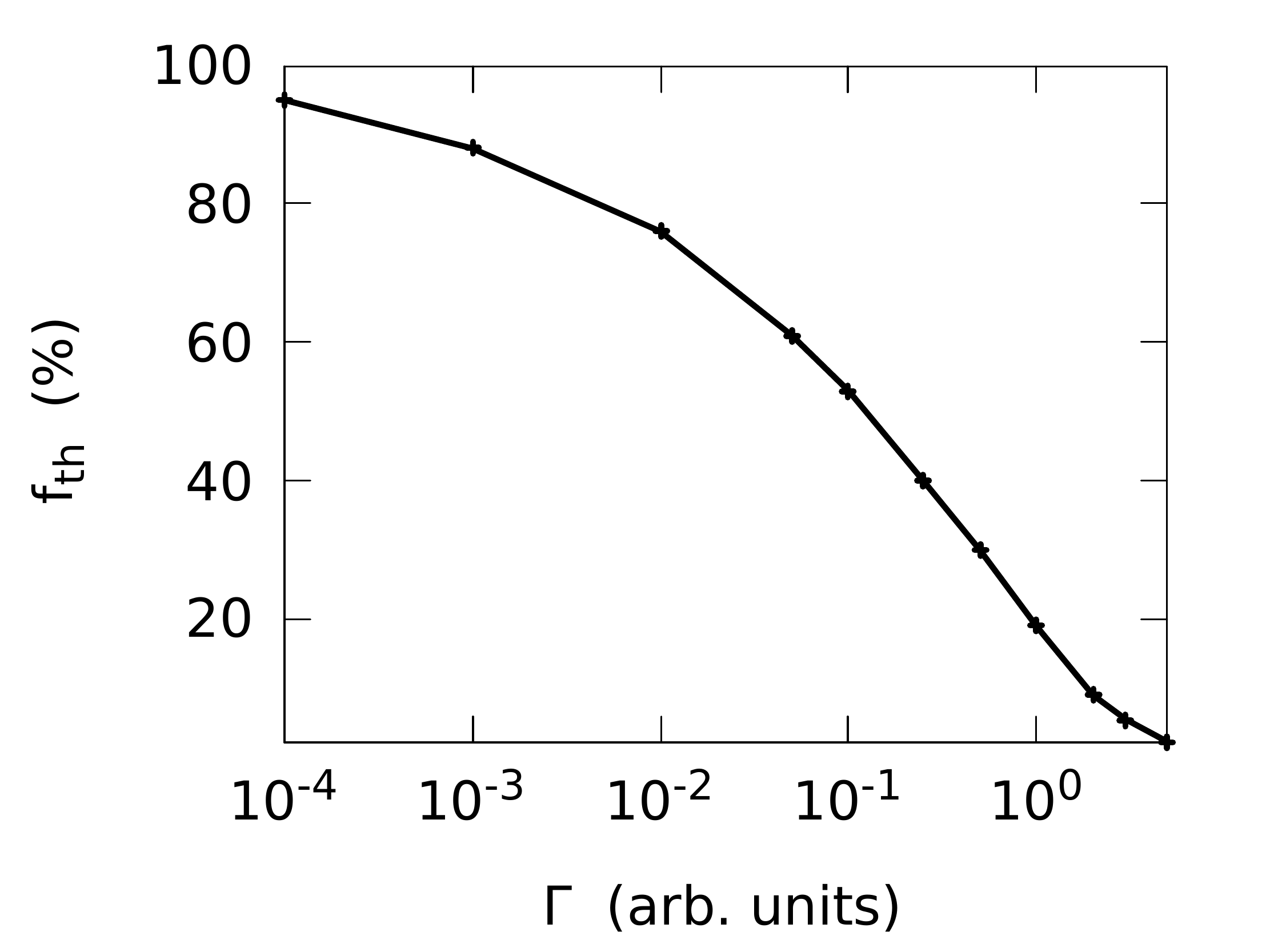}
	\caption{Theoretical transmitted fraction $f_\text{th}$ (in percentage) as a function of the dimensionless parameter $\Gamma$ defined by Eq. \ref{eq_k} in the case of a uniform, divergence-less ion beam aligned with the capillary axis.
\label{fig_theotrans_gamma}}
\end{figure}

Equation (\ref{eq_traj_dimless}) is solved numerically, using a standard 4$^\text{th}$ order Runge-Kutta method. For a given $\Gamma$ and for a given initial value  $\varrho_0$ the trajectory $\varrho(\zeta)$ is uniquely defined. If the trajectory is intercepted by the capillary wall, we reduce the initial value of $\varrho_0$, otherwise, $r_0$ is increased. Using the \href{https://en.wikipedia.org/wiki/Bisection_method}{bisection method}, a good estimate for the largest value of $\varrho_0^\text{max}$, for which the ion is still transmitted is  found. Assuming that the injected current density $j_0$  is uniform, one has $I_\text{in}=j_0 \pi R^2$ and $I_\text{out}=j_0 \pi (\varrho_0 R)^2$. The ratio 
\begin{equation}
\frac{I_\text{out}}{I_\text{in}}=(\varrho_0^\text{max}(\Gamma))^2=f_\text{th}(\Gamma)
\end{equation}
defines the theoretical transmitted fraction $f_\text{th}$ of the ion beam in the ideal case where the nano-capillary is a straight cylinder and the ion beam has zero divergence. The fraction $f_\text{th}$  depends only on the factor $\Gamma$ and  has  to be understood as an upper bound value, obtainable only in ideal conditions, i.e.  the injected beam is spatially uniform, mono-kinetic,  parallel and the capillary is perfectly cylindrical. In figure \ref{fig_traj_rho}, we show 3 trajectories for 3 different values of $r_0$  in the case of $\Gamma=1$. We see that only the ions that are injected at a distance less than $\varrho_0^{\text{max}}=0.434$  are transmitted, so that for $\Gamma=1$ the transmitted fraction  $f_\text{th}$ equals 18.6\%.
The theoretical transmitted fraction was calculated  for a large range of $\Gamma$ and is shown in Fig. \ref{fig_theotrans_gamma}. In the case of marco-capillaries, $\Gamma$ is typically well below  $10^{-4}$, yielding $f_\text{th}$ close to 100 \%. As expected, this indicates that the effect of the image charge force on the ion trajectory is negligible in macro-capillaries. However, for keV ions through nano-capillaries, $\Gamma$ is or the order of unit,   and the ions transmission is expected to be strongly reduced by the image charge force. For $\Gamma$ as large as 10, the transmitted fraction falls even below 1\%.
  
%
%In table (\ref{table_1}) we list the calculated  theoretical transmitted fractions ${(\rho_0^\text{max}/R)}^2$ for various capillary experiments found in the literature. 

\subsection{Comparison with experimental data}
\label{sec_highliting}
In  experimental works on nano-capillaries, authors usually give the transmitted ion rate 
$I_\text{out}$ through a large number of capillaries rather than the transmitted fraction $I_\text{out}/I_\text{in}$. Indeed,  the injected ion rate $I_\text{in}$ has to be deduced from the current density and  size of the beam, as well as transparency of the foil containing the capillaries, parameters which are difficult to obtain with sufficient precision. 
We nevertheless found three experimental studies which give the transmitted fraction at zero tilt angle. 
 
Surprisingly, when comparing the theoretical transmitted fractions $f_\text{th}      $ to experimental data found in \cite{Sahana_PRA_2006,Niko_PRA_2010,Niko_EPJ_2021}, we  noticed a significant disagreement.  In a recent work by Stolterfoht \textit{et al.} \cite{Niko_EPJ_2021},  a transmitted fraction of 50\% was observed for an 3-keV Ne$^{7+}$ ion beam through polyethylene terephthalate (PET) nano-capillaries. The latter have a length of $H=12$ $\mu$m and a diameter of $2R=230$ nm, yielding an aspect ratio of $H/(2R)\simeq 50$. Using for PET a dielectric constant of 3.0  \cite{Jin_2006},  one has for the parameter $\Gamma$ a value of 0.55. The theoretical transmitted fraction obtained for $\Gamma=0.55$ is 28\%, which is lower than the experimental one by almost a factor 2. In order to recover the experimental transmitted fraction of 50\%,
the parameter $\Gamma$  needs to  be reduced by a factor 4.

\begin{figure}[h!]
	\includegraphics[scale=0.38]{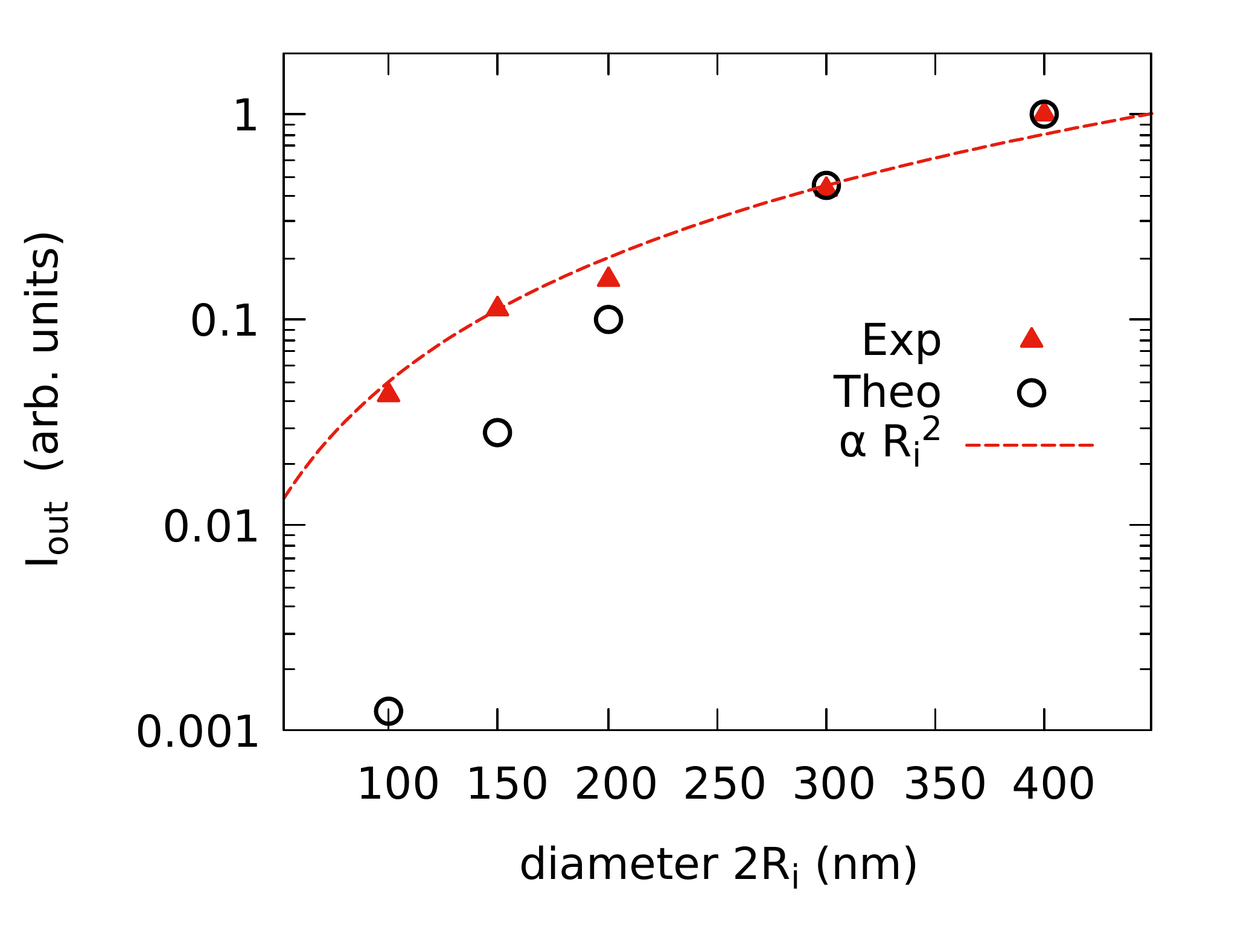}
	\caption{Comparison between observed transmitted current (red triangles)  extracted from \cite{Niko_PRA_2010} and our theoretically transmitted current (black circles) defined by Eq. \ref{eq_fth_current} as a function of the inner capillary diameter $2 R_i$. Dashed line shows the fit by a quadratic polynomial. 
\label{fig_table}}
\end{figure}

In 2010, Stolterfoht \textit{et al.} measured the transmission rate of PET nano-capillaries of various diameter $2R_i$, ranging from 100 to 400 nm \cite{Niko_PRA_2010}. They did not give the transmitted fraction for each diameter but rather the relative transmitted ion yield $I_\text{out}^\text{exp}(R_i)$. They found that at zero tilt angle, $I_\text{out}^\text{exp}(R_i)$ is grossly proportional to the section of the capillary, $$I_\text{out}^\text{exp} \propto  R_i^2.$$ This trend is shown in figure \ref{fig_table}, where the data (red triangles) were extracted from \cite{Niko_PRA_2010} and fitted by a $\alpha R_i^2$ function. Obviously, this analysis advocates that, for those measurements, the current density and spot size of the injected ion beam was kept constant. It follows that the injected current $I_\text{in}$ was  also proportional to the inlet section  $\pi R_i^2$ of the capillaries, $$I_\text{in} \propto  R_i^2,$$ so that  the transmitted fraction $I_\text{in}/I_\text{out}$ was found grossly independent from the capillary radius $R_i$. This is surprising because equation \ref{eq_k} tells us that $\Gamma$ scales as $R^{-3}$ and thus a strong dependence of the transmitted fraction on the capillary radius is expected, i.e. figure \ref{fig_theotrans_gamma}. We illustrate this discrepancy  in figure \ref{fig_table}, where we compare $I_\text{out}^\text{exp}$ extracted from \cite{Niko_PRA_2010}, to the theoretical values 
\begin{eqnarray}
I_\text{out}^\text{theo}(R) &=& f_\text{theo}(\Gamma(R))     \times I_\text{in}(R)  \notag \\
& \propto & f_\text{theo}(\Gamma(R)) \times  R^2 
\label{eq_fth_current}
\end{eqnarray}
for each capillary diameter used in \cite{Niko_PRA_2010}.  
For better comparison, the currents are normalized to unit for $2R_i=400$ nm.
While the measured transmitted current (red triangles) scales approximately as $R_i^2$ as indicated by the dashed red curve,  our model predicts that the transmitted current $I_\text{out}^\text{theo}$ decays much faster with decreasing capillary diameter.  In particular, for $2R_i = 100$ nm, the discrepancy is huge, the theoretical transmitted current $I_\text{out}^\text{theo}$ being 35 times smaller than the observed one.
This discrepancy makes us question the validity of expression (\ref{eq_im_force}) for the image charge force acting on a keV ion passing through PET nano-capillaries.

%
%\begin{table}[h]
%\begin{tabular}{c |c |c |c}
%\hline 
%\rule[-1ex]{0pt}{2.5ex} $2R_i$ (nm) & $I_\text{out}$  & $\alpha I_\text{out}/R_i^2$ & $f_\text{theo}$ \\ 
%\hline 
%\rule[-1ex]{0pt}{2.5ex} 100 & 3 & 0.35 & 0.01 \\ 
%\hline 
%\rule[-1ex]{0pt}{2.5ex} 150 & 8 & 0.41 & 0.1 \\ 
%\hline 
%\rule[-1ex]{0pt}{2.5ex} 200 & 11 & 0.32 & 0.2 \\ 
%\hline 
%\rule[-1ex]{0pt}{2.5ex} 300 & 30 & 0.41 & 0.4 \\ 
%\hline 
%\rule[-1ex]{0pt}{2.5ex} 400 & 70 & 0.5 & 0.5 \\ 
%\hline 
%\end{tabular}
%\caption{$I_\text{out}$ is the experimentally transmitted ion yield in arbitrary units, extracted from \cite{Niko_PRA_2010}, for PET capillaries of various diameter $2 R_i$. For each diameter, the normalized transmitted fraction  $\alpha I_\text{out}/R_i^2$ is compared to the theoretical transmitted fraction $f_\text{theo}$ of the capillary in question. The normalization factor $\alpha$ is chosen such that the  normalized transmitted fraction  equal 0.5 in the case of $2R_i=400$.
%\label{table_2} }
%\end{table}

In their  work of 2006 on highly ordered SiO$_2$ nano-capillaries, Sahana \textit{et al.} \cite{Sahana_PRA_2006} found that 20\% of the Ne$^{7+}$ ions, entering the
capillaries at zero tilt angle, are being transmitted. 
They used a 7 keV ion beam and the nano-capillaries had a length of $H=25$ $\mu$m and a diameter about $2 R\simeq 100$ nm, yielding a large aspect ratio of about $H/(2R) \simeq 250$. Taking for the dielectric constant of amorphous SiO$_2$  a value of 3.9 \cite{Kitamura_AO_2007}, we get using (\ref{eq_k}) for the factor $\Gamma$ a value of $14.9$.  For such a large $\Gamma$, our simulations find that only those ions which are injected within an initial distance $\rho\le  2$ nm from the symmetry axis are transmitted.  
The theoretical transmitted fraction $f_\text{theo}$ calculated with $\Gamma=14.9$ is  $0.2$\%, which is by a factor 100 lower than the experimental one.  In order to recover the experimental transmitted fraction of 20\%,
the scaling factor $\Gamma$ should be reduced here by a factor 14.

The observed transmitted fraction in those  three studies \cite{Sahana_PRA_2006,Niko_PRA_2010,Niko_EPJ_2021} are systematically significantly larger than those found theoretically in the ideal case. This seems to indicate that the image charge  force as given by Eq. \ref{eq_im_force} is not valid for keV ions through nano-capillaries. We remind that expression (\ref{eq_im_force}) was obtained  in the electrostatic limit with a static relative permittivity $\varepsilon_r$. 
The experimental data do not corroborate  the theoretical scaling of the transmitted fraction with the capillary radius. In particular, the influence of the image charge force in cylindrical capillaries with a diameter smaller the 200 nm seems to be much lower than what is predicted theoretically. 

\subsubsection*{Case of rhombic and rectangular capillaries}

Apart from cylindrical nano-capillarie, experimental studies were also done on muscovite mica nano-capillaries with rhombic (long axis 248 nm and  short axis 142 nm) and rectangular section (215  $\times$ 450 nm$^2$) \cite{Zhang_2012,Zhang_2017}. The observed beam-shaping of the transmitted ion was shown to be due to the image charge force, indicating clearly that the image charge force influences dominantly the ion trajectories.
In their simulations, the image charge force is a superposition of the image charge force for a plane dielectric surface for each of the 4 plane surfaces.
%, yielding the expression
%\begin{equation}
%\vec{F}_\text{im}(\vec{r})= \frac{q^2}{16 \pi \varepsilon_0} \frac{\varepsilon_r - 1}{\varepsilon_r + 1} \left( \frac{\vec{n}_1}{d_1^2}+\frac{\vec{n}_2}{ d_2^2}+\frac{\vec{n}_3}{ d_3^2}+\frac{\vec{n}_4}{ d_4^2} \right)  \quad, 
%\label{eq_im_force_rhombic}
%\end{equation}
%where $\vec{n}_i$ is the outgoing normal vector of one of the 4 edges of the rhombic or rectangular section and $d_i(\vec{r})$ the distance to the edge $i$. 
%This first order approach  accounts for the symmetry of the capillary section, but ignores the effect of the electric field generated by the polarization charges on the surface polarization itself. 
Interestingly, their simulations succeeded to reproduce the observed transmitted fraction of about 1\% for 7 keV Ne$^{7+}$ through the rhombic shaped muscovite mica nano-capillaries. However,  in the case of rectangular capillaries, the simulated transmission rate
and experimental values differ considerably (the experimental values are lower), but was attributed to size variations inside the channels. 

We conclude this section by noting that significant disagreements  were found between  measured and theoretical transmitted fractions in the case of cylindrical nano-capillaries made in PET and SiO$_2$. Better agreement was noticed in the case of rhombic shaped mica nano-capillaries, but was not further confirmed in nano-capillaries with rectangular sections. Where disagreement was fount, the experimental transmitted fraction was always significantly larger than the theoretical or simulated one.

\section{Velocity dependent  image charge force}
\label{sect_model_imforce}

%Because the ion beam is aligned with the capillary axis, the accumulated charge is expected to be uniformly distributed so that for  capillaries with large aspect ratio, the electric field inside the capillary can be considered negligible.

The disagreement highlighted in the previous section made us   question the validity of  expression (\ref{eq_im_force})  of the image charge force  in the case where  keV ions pass through a nanocapillary of large aspect ratio. 
There are indeed two effects that we ignored in the evaluation of the image charge force $ F_\text{im}(\rho)$. If the inner surface of the capillary is hit by ions, the accumulated charge may modify the dielectric response of the insulator. For example, in pre-charged PET samples, the trapped charge was found to partially inhibit the dipolar polarization, reducing by about 5\% the dielectric permittivity with respect to untreated PET \cite{Liu_2021}. However, except for \cite{Liu_2021}, which limited their study below the 1MHz domain,  no   other reference was found. 
So while the accumulated charge seem indeed to modify the dielectric response, the effect of excess charge on the dielectric response function is hardly documented in the literature and otherwise difficult to estimate by simple considerations. We decide thus for the present study to ignore the effect of the charge on the dielectric response. 

The other effect is that the ions are moving and the velocity of the ions influences the induced polarization at the interface. The dynamical effects in the image charge force were already investigated by \cite{Harris_1974} in the case of a metal plane interface. They found that dynamical corrections become non-negligible when the velocity $v$ of the moving charge approaches the Fermi velocity of electrons in metals. The reason is that the image  charge follows the mirror trajectory but is retarded by a distance $v t_r$ due to the finite response time, $t_r$ of the electron
gas. Similar findings are presented by  T\H{o}k\'{e}si \textit{et al.} \cite{Karoly_PRA_2001} where they studied  the dynamical correction of the image force in the case of a cylindrical conducting interface.  We will  in this section investigate the image charge force for a moving ion approaching a dielectric plane and cylindrical interface and evaluate in how far dynamical corrections modify the static image forge in the case of keV ions.

\subsection{Frequency dependent relative permittivity}

\begin{figure}[h!]
	\includegraphics[scale=0.36]{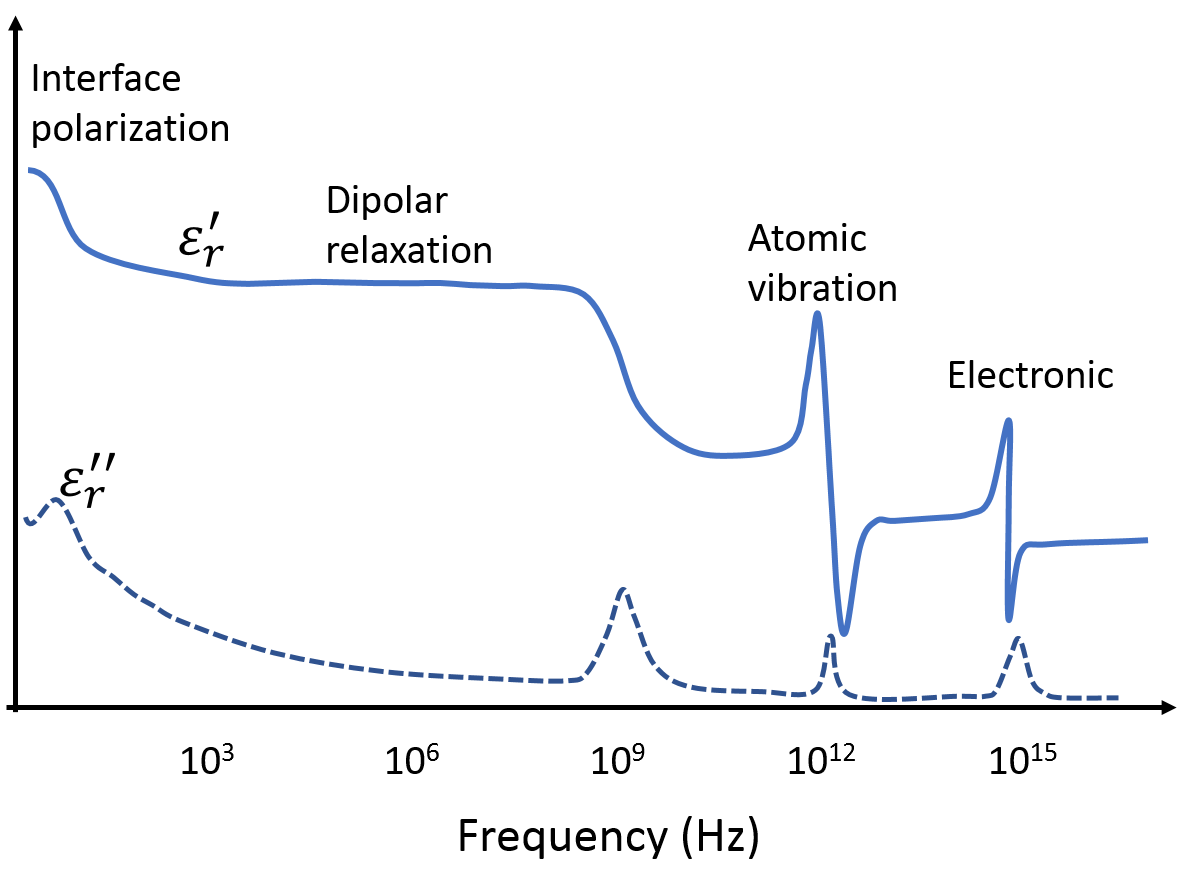}
	\caption{Arbitrary dielectric permittivity spectrum over a wide range of frequencies. The real $\varepsilon'_r$ and imaginary $\varepsilon''_r$ parts of permittivity are shown, and various processes are labeled: Interface polarization, dipolar relaxation,  atomic and electronic resonances at higher frequencies.}
\label{fig_permittivity_spectrum}
\end{figure}

The relative permittivity of bulk describes the ability to polarize a material subjected to an electrical field. In dielectrics, his polarization originates from a number of sources: electron cloud displaced relative to nucleus, relative displacement of atoms, alignment of dipoles in electric field and accumulation of charge carriers at the interface. These sources  have different response times so that the relative permittivity $\varepsilon_r \equiv \varepsilon_r(\omega)$ depends on the angular frequency $\omega$ of the external electric field.  
We assume  that the dependence of $\varepsilon_r$ on the wave vector $\vec{k}$ can be neglected if the spatial extension of the induced dipole moments is small  compared  to the wavelength $2\pi/k$, which is the case for frequencies of the electromagnetic field well below optical frequencies \cite{Petezlt}. The function $\varepsilon_r=\varepsilon'_r + i \varepsilon''_r$ is a complex quantity and the frequency-dependence of the real and imaginary part of $\varepsilon_r$  is illustrated in Fig. \ref{fig_permittivity_spectrum}. 
%In the low frequency domain, $f < 1$ kHz, space charge, dipolar, ionic and electronic polarization contribute all to the real part $\varepsilon_r'$ of the relative permittivity, while for frequencies in the optical domain $f >10^{14}$ Hz, only the electronic polarization contributes. The imaginary part of the relative permittivity  $\varepsilon_r''$ indicates the dissipation.

\begin{figure}[h!]
	\includegraphics[scale=0.4]{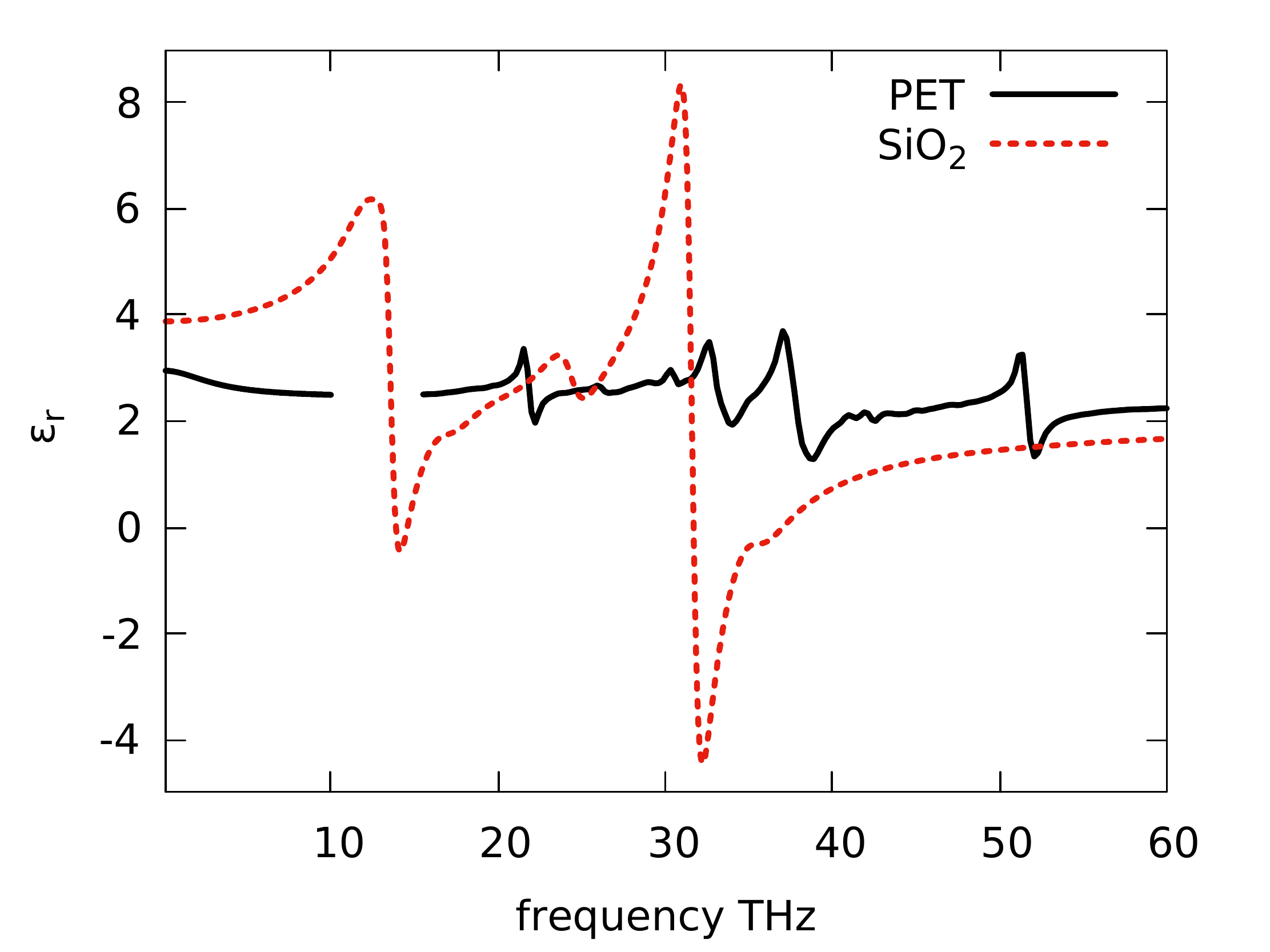}
	\caption{Real part of the dielectric response function of SiO$_2$ (red dashed line) and PET (black full line) extracted from \cite{Cataldo_2016} and \cite {Zhang_2020,Fedulova,YunSik_2006} respectively.}
\label{fig_er_SIO2_PET}
\end{figure}

For a keV ion moving along the interface, the electric field generated at a point of the interface may quickly evolve in time. If the response time of the dielectric function is too low, the patch of induced surface charges may trail the ion, which results in a reduction of the image charge force. Only polarization sources that have a sufficiently fast response time will contribute significantly to the image charge force. For a dielectric interface,  the image  force is thus expected to depend on the velocity of the ion \cite{Harris_1974,Karoly_PRA_2001}. Considering a keV ion having typically a velocity of about $10^5$ m/s and located 10 nm from the interface, the electric field generated by the ion at a point of the interface has a characteristic frequency of  $v/x_p=10$ THz, which is well in the infrared domain, where the dielectric response of insulators may vary significantly. In figure \ref{fig_er_SIO2_PET} we give the real part of the dielectric response of SiO$_2$ and PET insulators in the THz domain. Data were extracted from  \cite{Cataldo_2016,Zhang_2020,Fedulova} and show that the atomic dipole resonances responsible for a large part of the value of $\varepsilon_e$  are found above 10 THz.

\subsection{Velocity dependent image charge force}
\label{sec_velocity_dependent}

We will now evaluate the influence of the velocity of a moving  charge  on the image force. In a first step, we investigate the image force in the case of an infinite plane interface,  which yields for the velocity dependent image force a handy   expression, facilitating the discussion. Then, we check its effect on  in the case of a cylindrical interface.

\subsubsection{Case of an infinite plane interface }

Let us consider a homogeneous dielectric which fills the region $x < 0$. The region $x>0$ is empty. A projectile of charge $q$ is situated at $x=x_p>0$ and moves along the Oz coordinate with the velocity $v>0$.  The expression of image charge force acting on a moving ion has been detailed in appendix \ref{annex_1} and is given by
\begin{equation}
F_x(x_p,v) = \frac{q^2}{ 2 \pi \varepsilon_0 v^2}  \int_{-\infty}^\infty  \frac{d \omega}{2 \pi}|\omega| \frac{\varepsilon_r(\omega) -1 }{\varepsilon_r(\omega) + 1 }  K_1\left( \frac{2 x_p |\omega |}{v}\right)
\label{eq_im_force_plane_eps_omega}
\end{equation}
The image charge force depends on the velocity $v$ of the ion and on the frequency response of the dielectric function $\varepsilon_r(\omega)$.
For angular frequencies $\omega \gg v/(2 x_p)$,  the modified Bessel function $K_1(2\omega x_p/v )$ decreases  as $\exp(-2\omega x_p/v )$, so that the 
ratio  $v/(2 x_p)$ can be regarded as a cut-off angular frequency and only angular frequencies $\omega$ smaller  than   $v/(2 x_p)$ will contribute significantly to the image charge force. For example, 7 keV Ne$^{7+}$ ions (as used by Sahana \textit{et al.} \cite{Sahana_PRA_2006}) positioned $x_p=10$ nm from the interface yield a cut-off angular frequency of about $10^{13}$ radians Hz. The latter  is  close to the atomic polarization resonances, where the dielectric function varies quickly. Thus, for keV ions, the dielectric response function cannot be assumed constant and the  image charge force is expected to  depend significantly on the ion velocity.

For much lower ion  velocities, the cut-off frequency $v/(2 x_p)$ may fall below 1 Mhz.
In this low frequency range, the relative permittivity of insulators is usually constant, interfacial polarization apart. Taking  $\epsilon_r(\omega) =\varepsilon_r$ and integrating over $\omega$ yields the well-known image charge force for a plane vacuum-dielectric interface \cite{Jackson}. 
%\begin{eqnarray}
%F_x(x_p,v\rightarrow 0) &= &\frac{q^2}{ 2 \pi \varepsilon_0 v^2} \frac{\varepsilon_r -1 }{\varepsilon_r + 1 } \int_{-\infty}^\infty  \frac{d \omega}{2 \pi} |\omega|   K_1\left( \frac{2 x_p |\omega |}{v}\right) \notag \\
%&=&  \frac{q^2}{ 4 \pi \varepsilon_0 } \frac{\varepsilon_r -1 }{\varepsilon_r + 1 } \frac{1}{(2 x_p)^2}
%\label{eq_im_force_plane}
%\end{eqnarray}
\begin{equation}
F_x(x_p,v\rightarrow 0) = \frac{q^2}{ 4 \pi \varepsilon_0 } \frac{\varepsilon_r -1 }{\varepsilon_r + 1 } \frac{1}{(2 x_p)^2}
\label{eq_im_force_plane}
\end{equation}
Equation (\ref{eq_im_force_plane}) can thus be seen as the adiabatic limit of Eq. (\ref{eq_im_force_plane_eps_omega}), i.e. for  of vanishing  small ion velocities $v$.  

\begin{figure}[h!]
	\includegraphics[scale=0.4]{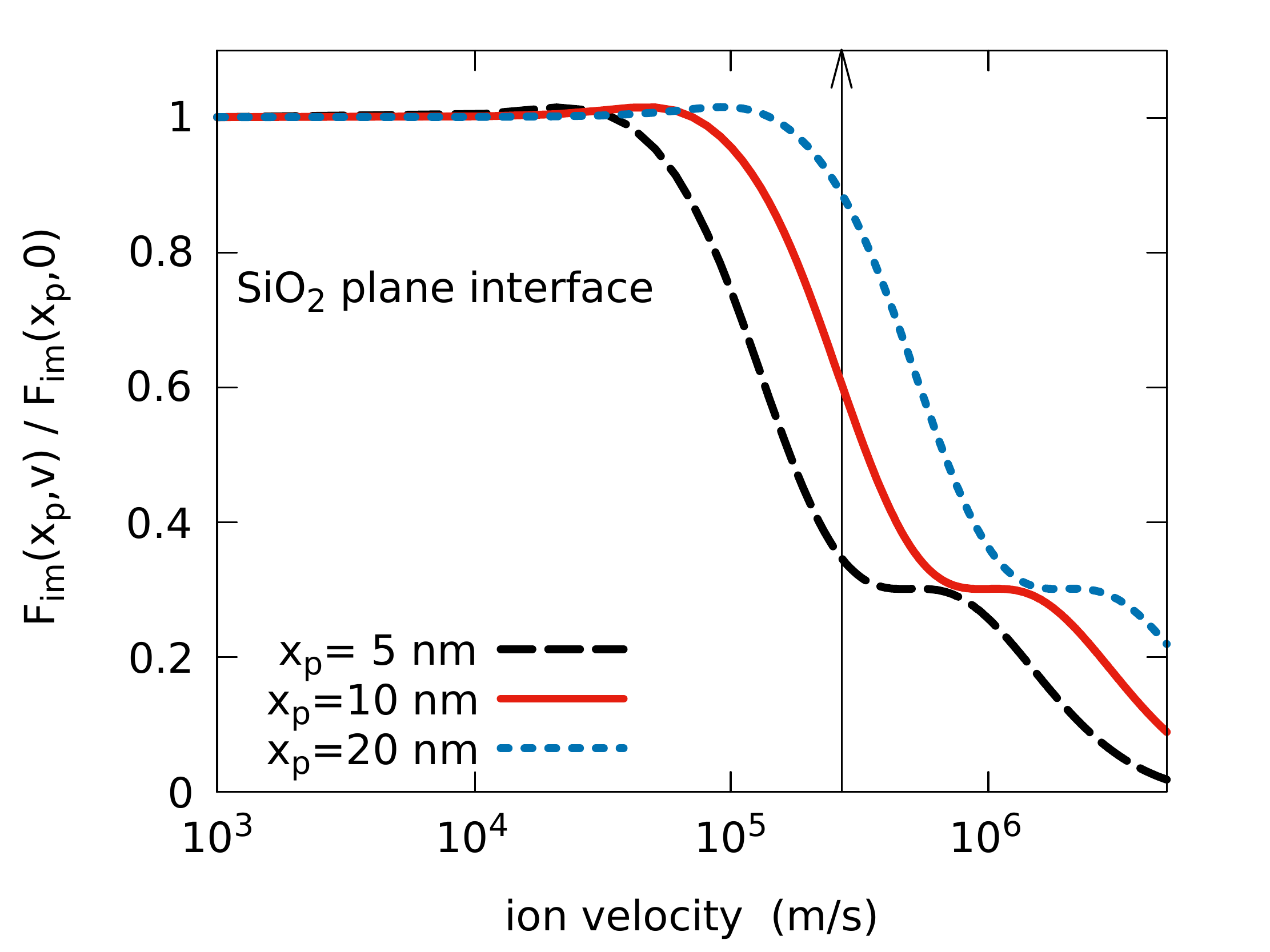}
	\caption{Normalized image charge force $F_\text{x}(x_p,v)/F_\text{x}(x_p,0)$ as a function of the ion velocity $v$, for 3 different distances $x_p$ from the SiO$_2$ interface, $x_p$ = 5,10 and 20 nm. The arrow indicates the velocity of 7 keV Ne$^{7+}$ ions as used by Sahana et al. \cite{Sahana_PRA_2006}.}
\label{fig_imforce_velovity}
\end{figure}

The dependence of the image charge force on the ion velocity is  shown in figure \ref{fig_imforce_velovity}. For better readability,  we normalize the latter with respect to the adiabatic ($v\rightarrow 0$)  image charge force (Eq. \ref{eq_im_force_plane}), which allows to highlight the influence of the ion velocity on the force.
The normalized force is shown for three different distances $x_p$ from the dielectric plane interface, namely $x_p$ = 5, 10 and 20 nm. Here, we used the dielectric response function $\varepsilon_r(\omega) $ of SiO$_2$ as shown in Fig. \ref{fig_er_SIO2_PET}, which has a large response in the THz domain, due to the ionic and atomic vibrations of the glassy network.  
For a distance $x_p=10$ nm from the interface, the normalized image charge force is reduced by 40\% for ion velocities larger than $10^5$ m/s. The reason is that for $v>10^5$ m/s,  atomic and ionic modes that usually lie in the THz range do not respond quickly enough to the time-varying  electric field. Only bound electrons respond quickly enough and contribute to the surface polarization.  For even larger velocities $v > 10^7$, bound electrons are not responding fast enough and the image force tends to zero. With this example we illustrate that the image charge force acting on a charge moving along a cylindrical interface is significantly reduced for velocities above a critical value, which depends on the frequency response of the dielectric function and distance $x_p$ from the dielectric interface. 

\subsubsection{Case of an infinite cylindrical interface}
\label{section_cyl_inter}
 
In the next step, we consider an infinite  cylindrical interface 
of radius $R$ having its symmetry axis along $Oz$. The interface  separates the inner vacuum from the outer insulating medium of dielectric permittivity $\varepsilon_r(\omega)$. The latter is assumed complex valued and depending on the frequency. 
Let a charge $q$ move in the xOz plane with the velocity $v \vec{u}_z$, so that its  time-dependent position is given by the vector $(\rho',0,v t)$, with $\rho'< R$.  The expression of the dynamical image force is similar to the static one (Eq. \ref{eq_im_force_exact})  but takes now explicitly into account the dielectric response and depends on the velocity $v$, 
\begin{equation}
F_\text{im}(\rho',v) = \frac{q^2}{16 \pi \epsilon_0}  \! \sum_{m=0}^\infty  \int_{0}^\infty \! \! \! \! dk  \left. \frac{\partial \text{I}_m^2(k \rho)}{\partial \rho} \right|_{\rho'} \! A_m(k R,\varepsilon_r( k v)) 
\label{eq_im_force_frequency_exact}
\end{equation}
%
%\begin{eqnarray}
%V_\text{im}(\rho,\phi,z,t) &=& \frac{q}{4 \pi \epsilon_0}  \sum_{m=0}^\infty  \cos(m \phi) \int_{-\infty}^\infty  e^{i \omega t} d\omega  \notag \\
%&  & \times \int_{0}^\infty dk  \cos(k z)   \text{I}_m(k \rho') \, \text{I}_m (k \rho)  \notag \\
%& & \times \, \, \chi_m(k R,\omega) \delta(|\omega|-k v)\;.\notag  \\
%\end{eqnarray}
%Integrating over $\omega$ and using the symmetries of the system one has
%
%\begin{eqnarray}
%V_\text{im}(\rho,\phi,z,t)&=& \frac{-q}{8 \pi \epsilon_0}  \sum_{m=-\infty}^\infty e^{i m \phi} \int_{0}^\infty \!  \!  \! dk \cos(k(z-vt))  \notag \\
%& & \times \, \text{I}_m(k \rho') \, \text{I}_m (k \rho) \, A_m\left(k R, \varepsilon_r (k v)\right) \;.
%\label{eq_im_pot_cyl} 
%\end{eqnarray}
%
The function $\text{I}_m ()$ and $\text{K}_m ()$ are the modified Bessel functions and  
$A_m()$ is a dimensionless function depending on   the dielectric  permittivity $\varepsilon_r(\omega)$, evaluated at the  pulsation $\omega  =k v$, 

%\lipsum[1]
%\begin{widetext}
\begin{eqnarray}
A_{m} \left( k R,\varepsilon_r(k v) \right) &= & \frac{\varepsilon_r(k v)-1}{  \left(\varepsilon_r(k v) \dfrac{\text{I}_m (k R)}{\text{K}_m (k R)} -\dfrac{\text{I}'_{m} (k R)}{\text{K}'_{m} (k R)}  \right)} \notag \\
\label{eq_chi_1}
\end{eqnarray}
%\end{widetext}
%\lipsum[1]
Here $\text{I}'_m $ and $\text{K}'_m $ are   the partial derivatives of the Bessel functions with respect to $\rho$. Note that the imaginary part of $A_m$ cancels so that $A_m$ is real valued. 
%The radial component of the electric field inside the cylinder due to the induced surface polarization charges is obtained by  differentiating  the potential $V_\text{im}$ (\ref{eq_im_pot_cyl})  with respect to $\rho$.
%Evaluating the electric field at the coordinates of the ion, $(\rho=\rho',\phi=0,z=vt)$, yields the image charge force acting on the moving ion,
%\begin{equation}
%F_\text{im}(\rho',v) = \frac{q^2}{16 \pi \epsilon_0}  \! \sum_{m}   \int_{0}^\infty \! \! \! \! dk  \left. \frac{\partial \text{I}_m^2(k \rho)}{\partial \rho} \right|_{\rho'} \! A_m(k R,\varepsilon_r( k v)) 
%\label{eq_im_force_frequency_exact}
%\end{equation}

We may highlight the special case of a perfectly conducting interface. For the latter, one has $\varepsilon_r(\omega)\rightarrow 1+i \infty $, and the functions $A_m$ tend to,  $A_{m} \rightarrow K_m(kR)/I_m(kR)$. The expression of the image charge force of a perfectly conducting interface becomes then independent of the ion velocity $v$ and can be integrated over $k$ and summed over $m$ to yield  \cite{Karoly_PRA_2001}
\begin{equation}
F^{\text{cond}}_\text{im}(\rho') = \frac{q^2}{4 \pi \epsilon_0 }  \frac{\rho' R}{(R^2-{\rho'}^2)^2} \quad.
\label{eq_imforce_conduct}
\end{equation}
Finallay note that in the adiabatic limit, $v \rightarrow 0$, the functions $A_m()$ are evaluated using for the relative permittivity its static limit,  $\varepsilon_r(k v\rightarrow 0 ) =\varepsilon_{r} $, yielding expression (\ref{eq_im_force_exact}).

\section{Influence of the ion velocity on the theoretical transmitted fraction}
\label{sec_numerics}
\subsection{Image force for keV ions in  SiO$_2$ and PET nano-capillaries}
\begin{figure}[h!]
\includegraphics[scale=0.38]{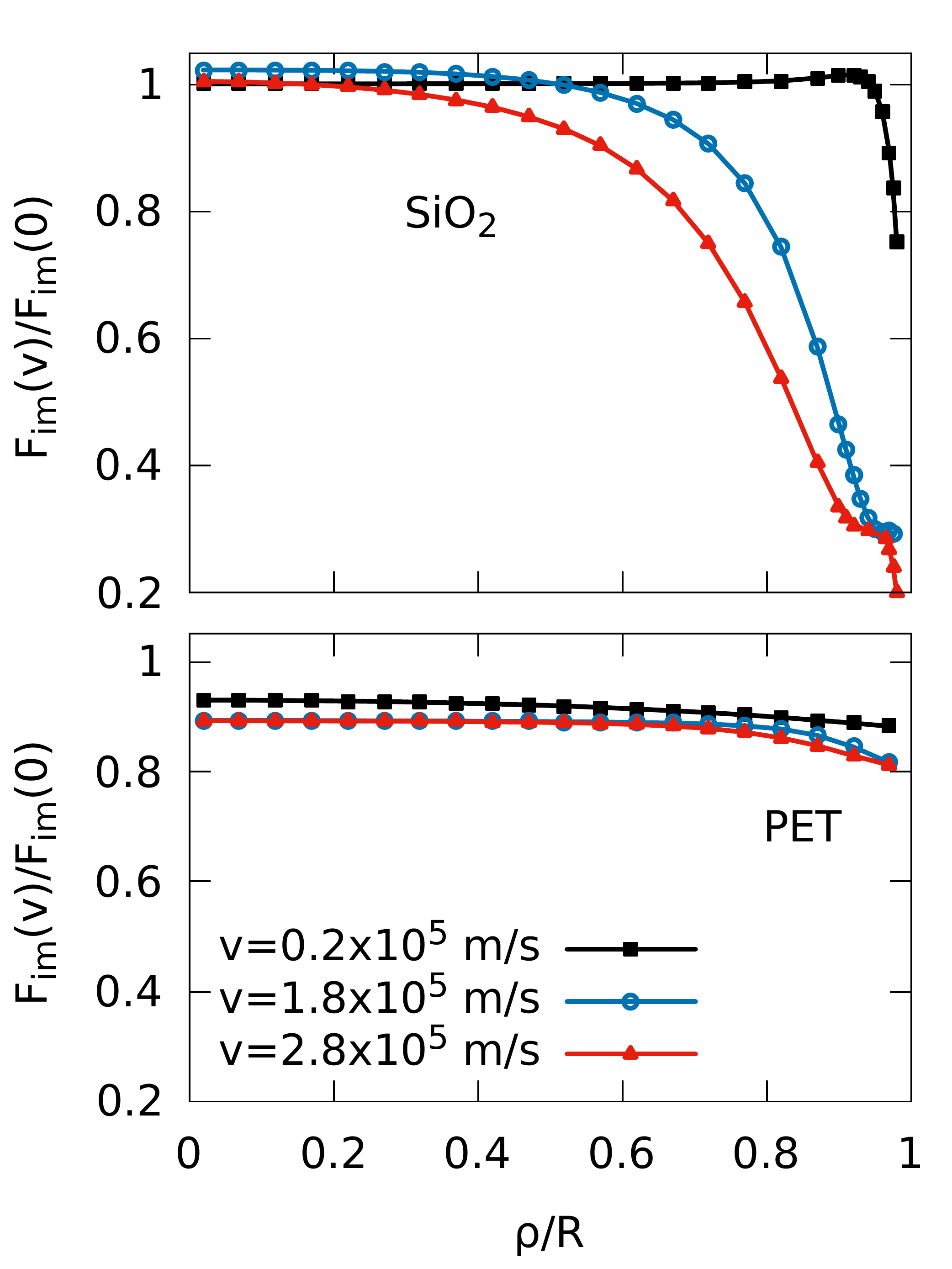}
	\caption{Normalized image charge force, $F_\text{im}(\rho,v)/F_\text{im}^0(\rho)$ of a moving charge inside a cylindrical dielectric interface of radius $R=50$ nm as a function of the dimensionless radial distance $\rho/R$, for three different velocities, namely  $v=2 \times 10^4$, $v=1.8 \times 10^5$, and $v=2.8 \times 10^5$ m/s. Upper panel, dielectric response function of SiO$_2$ as given in Fig. \ref{fig_er_SIO2_PET} was used. Bottom panel, dielectric response function of PET as given in Fig. \ref{fig_er_SIO2_PET} was used.}
\label{fig_imforce_cyl_rr_vv}
\end{figure}
We may now check in how far the velocities of keV ions reduce the image charge force for  those cases highlighted in section \ref{sec_highliting}, namely for  7 keV Ne$^{7+}$ ions through SiO$_2$ nanocapillaries  and  3 keV Ne$^{7+}$ ions through PET  nanocapillaries. 
The dielectric response functions of  SiO$_2$ nano-layers   and of  a thin samples of PET are pictured in Fig. \ref{fig_er_SIO2_PET}.  We compute for each case the image force as a function of $\rho$ and for three different velocities, namely $0.2 \times 10^5$ m/s, $1.8 \times 10^5$ m/s and $2.8 \times 10^5$ m/s. The two latter correspond to  Ne$^{7+}$ at 3 keV, as used by Stolterfoht \text{et al.} \cite{Niko_PRA_2010,Niko_EPJ_2021} and Ne$^{7+}$ at 7 keV ions, as used by Sahana  \textit{et al.} \cite{Sahana_PRA_2006}, respectively. The first smaller velocity,  is added for comparison. Expression (\ref{eq_im_force_frequency_exact}) is  evaluated numerically using MATHEMATICA \cite{MATHEMATICA} for the case of a  cylindrical interface of radius $R=50$ nm. The sum over $m$ was limited to the 100 first angular multipoles. The results are shown in Fig. \ref{fig_imforce_cyl_rr_vv}, where the dynamical image force $F_\text{im}(v)$ is normalized with respect to the static one, $F_\text{im}(v=0)$.
In the case of a SIO$_2$ interface  and 7 keV Ne$^{7+}$, (red triangles, upper panel, $v \simeq  2.8\times 10^5$ m/s), the dynamical image force is reduced by up to a factor 3 compared to the adiabatic image force when approaching the interface. Indeed, for such a velocity, the response time of the first Lorentz oscillator with an eigenfrequency about 12 THz is too low to contribute efficiently to image force. This results in a significant decrease  of the image force when the ion approaches the interface by about 23 nm or, in the present case of a cylindrical interface of  $R=50$ nm, for $\rho / R \ge  0.6$.  On contrary, for a 10 times smaller ion velocity (black squares), all the dipole oscillators follow the ion in phase and the dynamical corrections on the image force are negligible.

For a PET interface, the ratio $F_\text{im}(v)/F_\text{im}(0)$ for a moving charge of $1.8\times 10^5$ is about 0.88 and mainly independent on the radial coordinate $\rho$. This is due to the fact that the dynamical corrections to the image force come from  the finite response time of dipole alignment in PET, as indicated by the small decrease of $\varepsilon_r$ from 2 to 2.6 in the sub 3 THz domain, as shown in Fig \ref{fig_er_SIO2_PET} and reported by \cite{Fedulova}.
Also, the first resonance frequency of atomic dipoles in PET lies well above 20 THz and also has  a smaller strength than the first resonance in SiO$_2$. As a result, the dynamical image force  for charges moving with a velocity up to $2.8\times 10^5$ is only slightly affected when approaching the PET interface.
We deduce that for keV ions approaching a PET interface,  the dynamical corrections are small and the dielectric permittivity of PET is thus well approached by a dielectric constant $\varepsilon(f=3\text{THz})=2.6$.  

\subsection{New theoretical transmitted fraction}
We may now check to what extend the dynamical corrections to the static image force lift the discrepancy found between the experimentally observed and theoretically evaluated transmitted fraction, as described in section \ref{sec_highliting}. Again we solve equation \ref{eq_traj}, but this time using the image force as given in Fig. \ref{fig_imforce_cyl_rr_vv}. We look for the largest initial value of $\rho_0^\text{max}$ for with  7 keV Ne$^{7+}$ ions are transmitted through SiO$_2$ nanocapillaries, as those used by Sahana \textit{et al.}.  We get a theoretical transmitted fraction of 0.3\%, only slightly larger than the  0.2\% obtained previously using the static image force, but still well below the observed 20\% transmission. Similarly, for 3 keV Ne$^{7+}$ ions through PET nanocapillaries, as those used by  Stolterfoht \text{et al.} \cite{Niko_PRA_2010,Niko_EPJ_2021}, dynamical corrections increase  only negligibly the theoretical transmitted fraction. With respect to the one obtained with a static image force, $f_{\text{theo}}$ increases from 28\% to 28\%, which is still almost a factor 2 lower than the observed 50\% transmission.
We can thus affirm that even if the velocity of the keV ion is taken into account in the evaluation of the image force inside a dielectric cylinder, we are not able to explain the experimentally observed transmission fractions of ions through large aspect nano-capillaries, especially for inner diameters of about 100 nm.

\section{Conclusion}

We evaluated the theoretically  transmitted ion beam fraction  through straight insulating nano-capillaries by calculating the fraction of ion trajectories that were not intercepted by the capillary wall. The only force acting on the ions was the image charge force due to the induced polarization at the inner dielectric interface of a straight cylindrical capillary. For the calculation of the trajectories we used an CPU-friendly expression of the image force. We verified  that the latter  approaches well the exact image charge force in the adiabatic limit. The transmitted fraction was evaluated in the ideal case of a mono-kinetic divergence free beam injected at zero tilt angle. When comparing the  theoretically transmitted fraction to the three experimental data available in the literature, we found large discrepancies in the transmitted fractions,  especially for nano-capillaries with an inner diameter below 200 nm. 

%We expected thus that ion transmission through nano-capillaries is a observable sensitive enough to allow to refine the model of the image charge force of ions through nano-capillaries.
 
As the relative permittivity of a dielectric material depends on the frequency of the applied electric field,  the surface polarization induced by a charge moving along the interface depends on the its velocity. We evaluated thus the velocity depend image charge force for a plane and cylindrical SiO$_2$ and PET interface. We found that for keV ions having typically a velocity about $10^5$ m/s, the image charge force is controlled by the near infrared domain (THz) of the dielectric response function. For a SiO$_2$ interface, the velocity of the ions affects notably the image force when the ion is close to the interface. For PET interface, the dynamical effects of the image force are  negligible in the chosen keV range.  

We then checked in how far the theoretically transmitted  fractions are modified when using the velocity dependent image charge force instead of the static (adiabatic) one. The calculations were performed for 7 keV and 3 keV Ne$^{7+}$ ions. We found that the newly calculated transmitted fractions are only slightly larger than those found previously in the case of  adiabatic limit and still disagree with the experimental results. We conclude thus that a velocity dependent image charge force is not able to  lift the highlighted disagreement. And because the electric field inside the capillary due to an uniformly charged interface can be neglected for large aspect ratio capillaries, there is still no  convincing explanation for why the theoretical transmitted fraction was found well below the observed one, especially for capillaries with inner diameters below 200 nm. 
With this work we hope to motivate new experimental work on the ion transmission through insulating nano-capillaries of different sections. We think that new data will guide authors to  model more accurately the image force of keV ions trough capillaries, shedding new light on the phenomenon.

\section{Appendix}
\subsection{Electric field of the deposited charges vs image force  on ion trajectories}

\label{appendix_radial_electric}

\begin{figure}[h!]
	\includegraphics[scale=0.35]{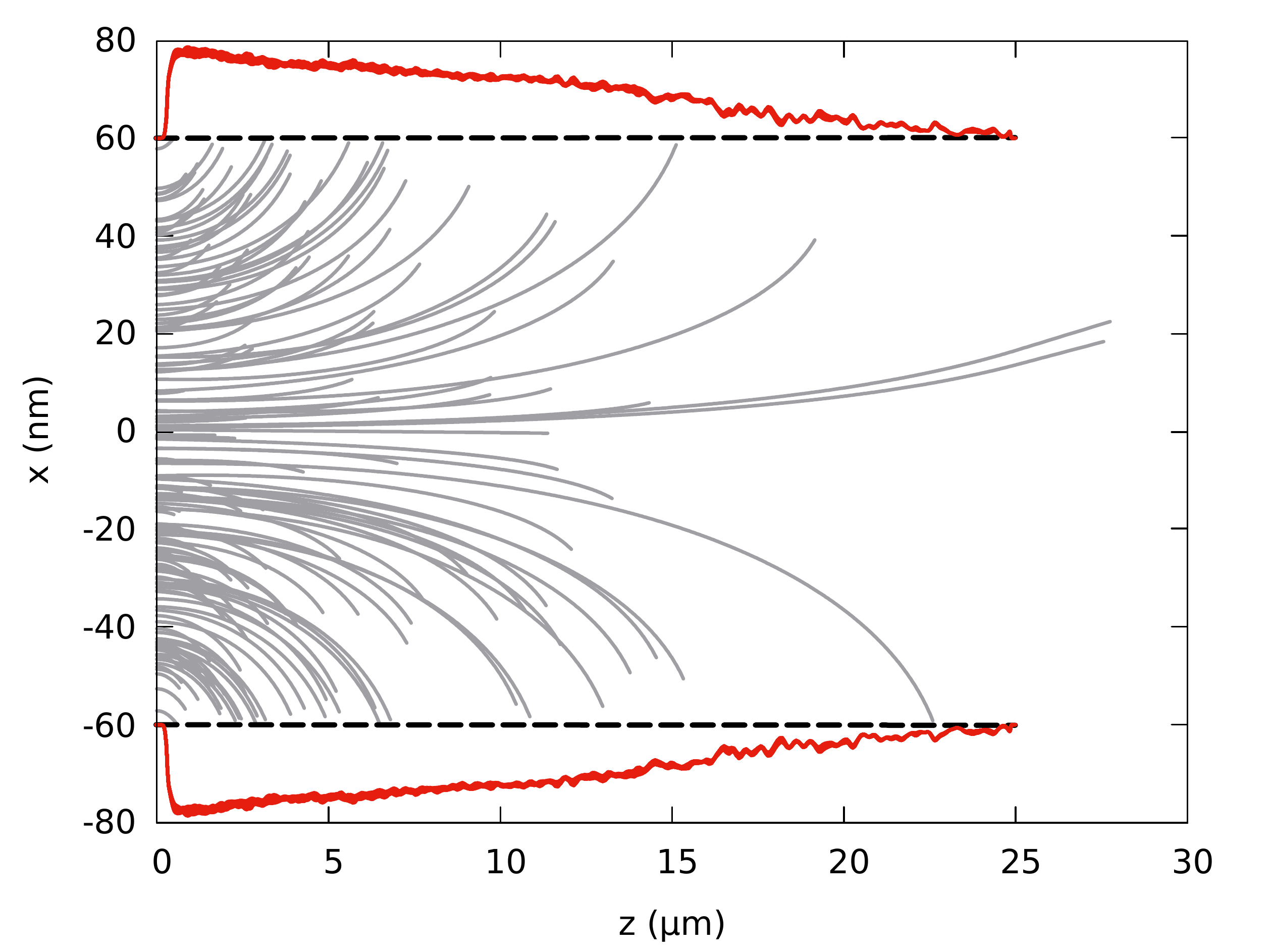}
	\caption{Simulated ion trajectories (gray curves) through a 25 $\mu$m long nanocapillary of inner radius of 60 nm, as indicated by the black dashed line. The red curves indicate the accumulated charge in arbitrary units at the inner interface. 
\label{fig_traj_sio2}}
\end{figure}
We simulated the trajectories of  7 keV Ne$^{7+}$ ions through an insulating SiO$_2$ nano-capillary embedded in semi-conducting silicon. The inner radius of the capillary tube  is 60 nm, the outer radius is 150 nm and the length is 25 $mu$m.  The forces acting on the ions are the image force expressed by Eq. \ref{eq_im_force} and the Coulomb force due to accumulated charges at the inner capillary surface. The beam axis is aligned with the capillary axis and the injected beam intensity is 10 ions per second, thus about $10^{-17}$ A.  The outer surface is grounded and  accounts for the grounded semi-conducting  Si layer. The bulk conductivity $\kappa$  of SiO$_2$ is field dependent and given by the non-linear Poole-Frenkel relation \cite{Frenkel}
\begin{equation}
\kappa(E)=\kappa_0 \exp\left(\sqrt{E/E_c}\right) \quad.
\end{equation}
The field independent bulk conductivity  $\kappa_0$ is set to $10^{-15} $ S/m \cite{Srivastava} and the critical field $E_c$ is set be $10^{7}$ V/m, which is the typical range where the conductivity SiO$_2$ becomes non-linear \cite{Dash_2018}. We show in Fig. \ref{fig_traj_sio2} the computed  ion trajectories. The accumulated  charge in the capillary wall is shown by the red curves. It has reached its asymptotic distribution, where the deposited charge per unit time is balanced by the leakage current. The electric field inside the capillary wall is at that time about 30 MV/m.
The transmitted beam fraction as a function of time is shown in Fig. \ref{fig_trans_sio2}, where it is compared to the case where the electric field due to the deposited charges is removed, so as to put into evidence the effect of the deposited charge on the transmission.

\subsection{Image charge at plane interface}
\label{annex_1}
Let us consider a homogeneous dielectric of relative permittivity $\varepsilon_r$  which fills the region $x < 0$. The region $x>0$ is empty. A projectile of charge $q$ is situated at $x_p>0$, $y_p=0$ and moves along the Oz coordinate with the velocity $v>0$, $z_p=v t$.  The x-component of the electric field at the interface $x=0$ generated by the moving charge  at position $\vec{r}_p(t)=(x_p,0,vt)$ reads
\begin{equation}
E^q_x(x_p,y,z,t)=\frac{q}{4 \pi \epsilon_0} \frac{-x_p}{(x_p^2+y^2+(z-v t)^2)^{2/3}} \quad.
\end{equation}
Because of the discontinuous jump of the relative permittivity $\varepsilon_r$ at the interface $x=0$,  the electric field $E_x^q$ induces at the interface surface charge density  $\sigma$. The electric filed  generated by the surface charges $\sigma$ at the dielectric side  $x=0^-$ is $-\frac{\sigma}{2 \varepsilon_0}$ but $+\frac{\sigma}{2 \varepsilon_0}$ at $x=0^+$, so that the total electric field at each side  reads
\begin{equation}
E_x^-= E^q_x - \frac{\sigma}{2 \varepsilon_0}  \quad, \quad x=0^-
\end{equation}
while in the empty region side, $x=0^+$, it reads 
\begin{equation}
E_x^+=E^q_x  + \frac{\sigma}{2 \varepsilon_0}  \quad,\quad x=0^+
\end{equation}
In the absence of free surface charges,  the x-component of the displacement field $D_x$ is continuous across the interface. The relation between $\vec{D}=\varepsilon \vec{E}$ is non-local in time and position and is more comfortably  used in the frequency  and  wave-number domain. 
The latter relation is more convenient when expressed in the Fourier space, where it simplifies to
\begin{equation}
\vec{\tilde{D}}(\vec{k},w) =  \tilde{\varepsilon}(\vec{k},\omega) \vec{\tilde{E}}(\vec{q},\omega) \quad, 
\label{E-Drelat}
\end{equation}

Let $\tilde{E}_x^q$ be the Fourier transform in the frequency domain  of the  electric field $E^q_x$,
\begin{eqnarray}
\tilde{E}_x^q(\omega,x_p,y,z,v) &=& \frac{-q}{4 \pi \epsilon_0}\int_{-\infty}^\infty  \frac{x_p e^{i w t} dt }{(x_p^2 + y^2 +(z-v t)^2)^{3/2)}} \notag \\
&=& \frac{-q\, e^{i w z/v}}{2 \pi \epsilon_0  } \frac{|\omega|}{v^2}\underbrace{\frac{  K_1\left( \frac{|\omega | \sqrt{x_p^2+y^2}}{v}\right)}{ \sqrt{x_p^2+y^2}} }_{ G(\omega,x_p,y,v)} \notag \\
\end{eqnarray}
where $K_1()$ is the modified Bessel function of order 1.

In the following we assume that the dielectric function is local in space so that 
$\varepsilon_r(\omega,\vec{k}) \equiv \varepsilon_r(\omega)$. This assumption is valid for $|k|\ll \pi/a$, where $a\simeq 0.5$ nm is the typical length of the molecules responsible for the polarization. 
Using (\ref{E-Drelat}) one has at both sides of the interface $x=0$, 
%\begin{widetext}
\begin{eqnarray}
\tilde{D}_x^- (\omega,y,z,v) &=&  \varepsilon_r(\omega) \left( \frac{\tilde{\sigma}(\omega,y,z,v)}{2}+\varepsilon_0 \tilde{E}_x()  \right) \\
\tilde{D}_x^+ (\omega,y,z,v) &=&  \left(-\frac{\tilde{\sigma}(\omega,y,z,v)}{2}+\varepsilon_0 \tilde{E}_x() \right) 
\end{eqnarray}
%
%which leads to
%\begin{eqnarray}
%D_x^- (\omega,k_z,y,v) &=&  \varepsilon_r(\omega,k_z) \left( \frac{1}{2}\sigma(\omega,k_z,y,v)+\frac{q}{2 \pi  }  \frac{|\omega|}{v^2} G(\omega,x_p,y,v)\delta(\frac{\omega}{v}-k_z  \right) \notag \\
%D_x^+ (\omega,k_z,y,v) &=&  \left(-\frac{1}{2}\sigma(\omega,k_z,y,v)+  \frac{q}{2 \pi  }\frac{|\omega|}{v^2} G(\omega,x_p,y,v)\delta(\frac{\omega}{v}-k_z) \right) \notag 
%\end{eqnarray}
Using  the interface condition $$\tilde{D}_x^-(\omega,x,y,v) = \tilde{D}_x^+ (\omega,x,y,v)$$ one obtains the expression for $\sigma(\omega,y,z,v)$
%\end{widetext}
%
%Using the inverse Fourier transform  $\sigma(\omega,z)=\int dk_z \sigma(\omega,k_z)e^{-i k_z z} $  yields the induced polarization charges at the interface,
\begin{equation}
 \tilde{\sigma}(\omega,y,z,v)= \frac{-q   e^{i w z/v}}{\pi }  \frac{\varepsilon_r(\omega) -1 }{\varepsilon_r(\omega) + 1 } \frac{|\omega|}{v^2} G(\omega,x_p,y,v)
\end{equation}

%Assuming that $| \varepsilon''_r(\omega)| \ll | \varepsilon'_r(\omega)|$ we get
%
%\begin{eqnarray}
%\sigma(\omega,y,z,v) &=& \left( \frac{\varepsilon'_r(\omega) -1 }{\varepsilon'_r(\omega) + 1 } \cos(wz/v) - \frac{2 \varepsilon''_r(\omega)}{(\varepsilon'_r(\omega) + 1)^2 } \sin(wz/v)\right) \notag \\
%& & \times  \frac{q   }{\pi } \frac{|\omega|}{ v^2} G(\omega,x_p,y,v) \notag 
%\end{eqnarray}
% 

The image charge force acting on the projectile is eventually,
\begin{equation}
F_x(x_p) = \frac{q}{4 \pi \varepsilon_0 }\int_{-\infty}^\infty \frac{d \omega}{2 \pi} \int dy \int \frac{ e^{-i\omega t} x_p \tilde{\sigma}(\omega,y,z,v) d z }{(x_p^2 + y^2 +(z-vt)^2)^{3/2}}
\end{equation}
Integration over $z$ gives 
\begin{equation}
F_x(x_p) = \frac{-q^2}{ 2 \pi^2 \varepsilon_0 v^3}  \int_{-\infty}^\infty  \frac{d \omega}{2 \pi} |\omega|^2 \frac{\varepsilon_r(\omega) -1 }{\varepsilon_r(\omega) + 1 }  \int_{-\infty}^\infty dy  G(\omega,x_p,y,v)^2 
\end{equation}
Noting that 
\begin{equation}
\int_{-\infty}^\infty dy  G(\omega,x_p,y,v)^2 = \pi \frac{v}{|\omega|}K_1\left( \frac{2 x_p |\omega |}{v}\right)
\end{equation}
one has
\begin{equation}
F_x(x_p) = \frac{-q^2}{ 2 \pi \varepsilon_0 v^2}  \int_{-\infty}^\infty  \frac{d \omega}{2 \pi}|\omega| \frac{\varepsilon_r(\omega) -1 }{\varepsilon_r(\omega) + 1 }  K_1\left( \frac{2 x_p |\omega |}{v}\right)
\end{equation}

\end{document}